# New detectors of the Experimental complex NEVOD for multicomponent EAS detection

I.I.Yashin[1], N.S. Barbashina[1], A.A. Borisov[1,2], A. Chiavassa[1,3], R.M. Fakhrutdinov[1,3], D.M. Gromushkin[1], S.S. Khokhlov[1,] R.P. Kokoulin[1], A.S. Kozhin[1,2], A.A. Petrukhin[1], I.A.Shulzhenko[1], Yu.V. Stenkin[1,4], E.A. Zadeba[1]

[1]*National Research Nuclear University MEPhI (Moscow Engineering Physics Institute), 115409, Moscow, Russia*
[2]*RF SSC Institute of High Energy Physics, 142281, Protvino, Russia*
[3]*Dipartimento di Fisica dell' Universita di Torino et INFN, 10125, Torino, Italy*
[4]*Institute for Nuclear Research of RAS, 117312, Moscow, Russia*

Experimental complex (EC) NEVOD includes a number of unique experimental facilities for studies of main components of cosmic rays on the Earth's surface. The complex is used for the basic research of CR flux characteristics and their interactions in the energy range $10^{15}$ - $10^{19}$ eV, and for applied investigations directed to the development of methods of the muon diagnostics of the atmosphere and the Earth's magnetosphere and near-terrestrial space. To extend the experimental capabilities and raising the status of the installation to the Mega Science level, nowadays new large-scale detectors: array for the EAS registration - NEVOD-EAS, detector of atmospheric neutrons - URAN, and large-area coordinate-tracking detector – TREK, are being deployed around EC NEVOD. The description of new detectors and a common trigger system to ensure the joint operation together with other detectors of EC NEVOD are presented.

## 1. INTRODUCTION

Experimental complex NEVOD [1] is located in the campus of MEPhI (Moscow) and is the only one in the world that allows to conduct the basic (particle physics and astrophysics) and applied (monitoring and forecasting of the state of near-terrestrial space) studies using cosmic rays (CR) on the Earth's surface in the entire range of zenith angles (0 to 180 degrees) and in a record energy range of CR primary particles (1 - $10^{10}$ GeV).

During 2002 - 2007 experimental series at the complex NEVOD, a new approach to the study of primary cosmic rays, based on the new EAS variable - local muon density spectra (LMDS) has been developed [2, 3]. The measurements were conducted with main detectors of the complex – the large volume Cherenkov water detector NEVOD and the large area coordinate detector DECOR. The analysis of the measured LMDS, revealed a significant excess of multi-muon events generated by CR with energies $10^{15}$-$10^{18}$ eV in comparison with calculated on the basis of modern models of hadron interactions (even at the assumption about the pure iron composition of the CR spectrum) [4]. This problem, named "the muon puzzle", was confirmed by other experiments [5, 6], in particular, the Pierre Auger Observatory [7]. Mainly, the further development of the Experimental complex NEVOD was targeted to the solution of this phenomenon.

At present, for the investigation of the muon puzzle a number of new detectors which will provide a multi-component registration of EAS are being designed and created around the complex.

## 2. EXISTING EXPERIMENTAL COMPLEX NEVOD

The complex NEVOD [8] includes several unique detection systems for the registration of the main components of cosmic rays at ground level: the large volume Cherenkov water detector (CWD) [9]; the vertically deployed around the CWD the coordinate tracking detector DECOR [10, 11] with high spatial ~ 1 cm and angular ~ 1° resolution for the detection of multi-particle events at large zenith angles, up to the horizon; the muon hodoscope URAGAN with a total detection area of 46 m$^2$ for continuous registration of muon flux in the range of zenith angles from 0 till 80 degrees [12, 13]; the system of calibration telescopes (SCT), which allows to calibrate detection system responses and detect electromagnetic and muon components of EAS [14]; the prototype of the detector for registration of atmospheric neutrons PRISMA-32 [15].

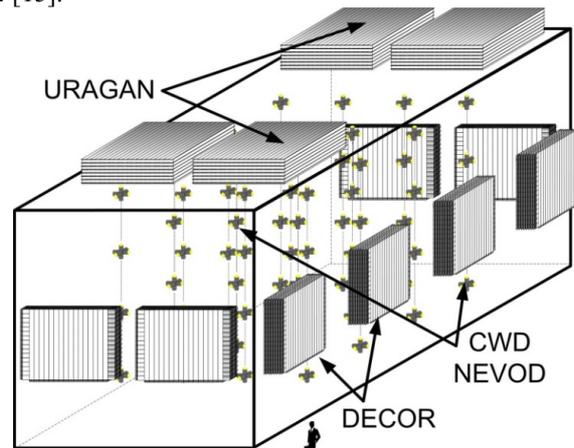

Figure 1: Scheme of the NEVOD-DECOR complex.

The layout of the NEVOD-DECOR complex is shown in figure 1. The basis of the complex is Cherenkov water detector with volume 2000 m$^2$ of purified water (9×9×26 m$^3$). CWD NEVOD is a multi-purpose detector, designed for the registration of all the main components of CR on the Earth's surface. The detection system of the NEVOD is represented by a regular spatial lattice of the quasispherical measuring modules (QSM). The QSM was specially designed to provide isotropic sensitivity [16]. Each module consists of six photomultipliers FEU-200 (EKRAN company [17], Russia) with flat cathodes of 15 cm in diameter, placed in an aluminum housing and oriented along the orthogonal coordinates axes (see figure 2). The sum of the squares of the amplitudes of the signals of triggered PMTs of single QSM is independent on the direction of Cherenkov light arrival. QSMs are arranged in strings with a step of 2.5 m along water tank and 2.0 m across it and over the depth. Therefore, the NEVOD measuring system has 4π sensitivity.

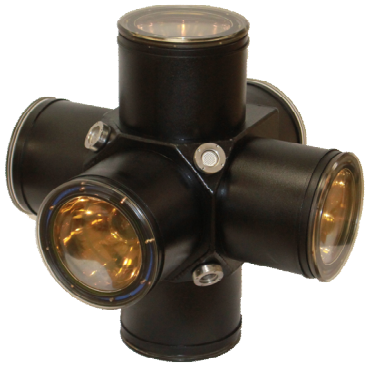

Figure 2: Quasi-spherical measuring module of the CWD NEVOD.

Two precise coordinate detector systems are deployed around the CWD – DECOR and URAGAN.

The coordinate detector DECOR represents 8 supermodules (SM, total area is ~ 70 m$^2$) vertically arranged in the galleries surrounding the water tank. Each SM consists of eight vertical planes of streamer tube chambers with external strip read-out system [18]. SM's angular and spatial resolutions are ~ 1° and 1 cm respectively. The main task of NEVOD-DECOR complex is the study of the muon component in wide ranges of multiplicities and zenith angles (up to the horizon).

Four SMs of the URAGAN hodoscope with total area about 46 m$^2$ are horizontally deployed above the NEVOD water tank. Muon hodoscope URAGAN is used for the detection and analysis of the study of angular variations of muon flux caused by various processes in the heliosphere, magnetosphere and the atmosphere of the Earth.

## 3. TREK SETUP

To solve the muon puzzle problem, it was proposed to measure simultaneously characteristics of muon bundles and their energy deposits in the CWD which represents after a deep modernization, a large volume homogenous Cherenkov water calorimeter (CWC) [19]. However, DECOR supermodules do not overlap the whole aperture of the Cherenkov water detector and do not exclude the passage of part of muon bundle within of the gaps between SM, and also, coordinate detector does not make it possible to resolve two tracks in the space at a distance of less than 3 cm. The new detector TREK [20] will completely cover the side aperture of the CWD NEVOD and will improve the resolution of close tracks byan order of magnitude. The TREK facilities represent vertical planes of the drift chambers (IHEP, Protvino) [21] with *X-Y* orientation, placed on the outside wall of the building of the complex (see figure 3). Each drift chamber has a large effective area (1.85 m$^2$), good spatial and angular resolution with a small number of measuring channels.

The detector will be operated as a part of the experimental complex NEVOD, in particular, together with Cherenkov water detector and DECOR. Each coordinate plane includes 132 chambers. The effective area of the detector is about 270 m$^2$.

At present the coordinate tracking unit on the drift chambers (CTUDC) as the prototype of full-scale TREK facility is mounted on the opposite sides of CWD in the short galleries, one floor above the DECOR supermodules location. It consists of two coordinate planes containing 8 drift chambers. One of the main goals of this setup is to elaborate the conditions of the joint operation with CWD and DECOR.

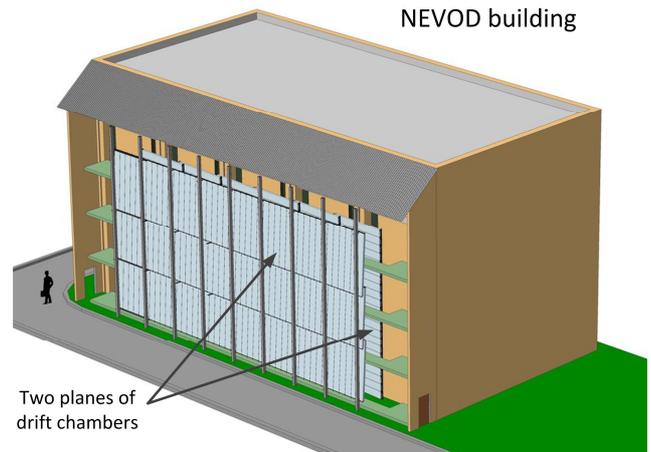

Figure 3: Scheme of the TREK facility.

## 4. NEVOD-EAS ARRAY

Another detector, which began operation in the EC NEVOD, is the traditional EAS detector array placed on the roofs of laboratory buildings surrounding the NEVOD on the territory of the MEPhI campus [22, 23]. The reason for the deployment of a traditional EAS detector is relatively low resolution for the PCR energy reconstruction via LMDS method: $\sigma_{lgE} \approx 0.4$. Measuring system of the NEVOD-EAS is formed on a cluster principle. Each cluster includes 16 scintillation counter integrated in four detecting stations and combined by a common Local Post (LP) of the detector array DAQ. For detection of EAS electromagnetic components, scintillation counters previously operated in the KASCADE-Grande detector are used [24, 25]. Picture of the scintillation detector and the scheme of the NEVOD-EAS detector station are shown in figures 4 and 5.

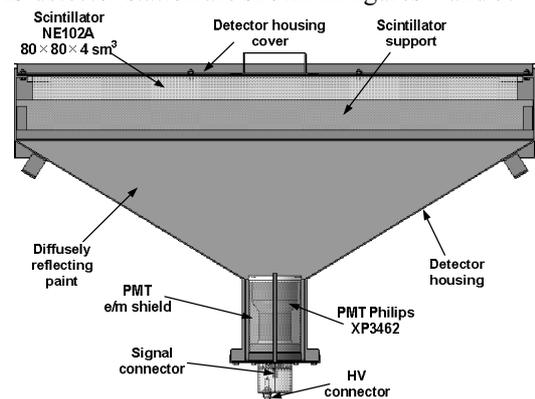

Figure 4: The scheme of the NEVOD-EAS counter.

NEVOD-EAS setup for registration of extensive air showers being created on the basis of the experimental complex NEVOD will allow determination of the size, position of the axis and the arrival direction of EAS with energies $10^{15}$-$10^{17}$ eV and will give the opportunity to test a method of reconstruction of the characteristics of PCR on the basis of muon bundle registration technique by means of the DECOR detector. New data obtained with NEVOD-EAS setup will allow to narrow the energy range of CR particles responsible for generation of muon bundles with certain multiplicity arriving at various zenith angles.

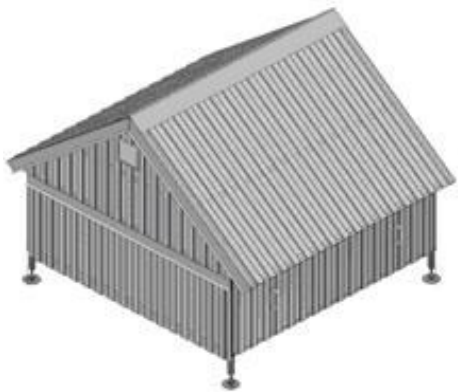

Figure 5: NEVOD-EAS detector station housing.

## 5. URAN SETUP

Currently, around NEVOD the system for registration of atmospheric neutrons – URAN – is being deployed [26, 27]. The first stage of the URAN array includes 72 *en*-detectors for simultaneous registration of electron-photon (*e*) and neutron (*n*) EAS components. The *en*-detectors are combined into independent cluster structures of 12 detectors. Clusters are located on two roofs of the laboratory buildings (3 clusters on each roof). The layout of the URAN setup is shown in figure 6. The typical distance between the detectors is 4–5 meters. The total area of the setup is $\sim 10^3$ m$^2$.

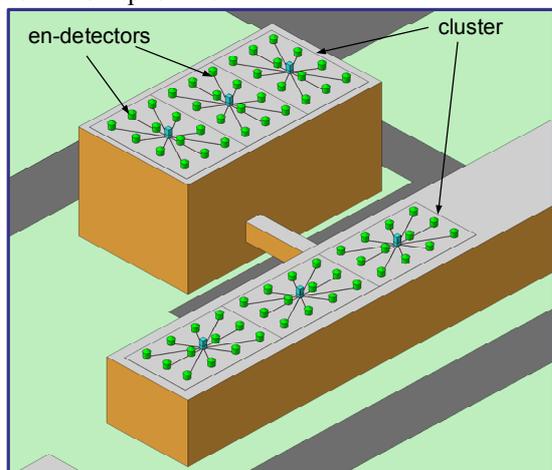

Figure 6: The layout of URAN setup clusters.

To ensure the operation of the detectors and primary cluster data processing, cluster's Local Posts (LP) is used. The structure of the URAN cluster is shown in figure 7. Local Post provides selection of events according to the cluster trigger conditions, digitization of amplitude information and data transmission to the Central DAQ Post via optical link. LP includes the crate with two 12-channel boards of amplitude analysis and controller, summator-multiplexer and its controller, and power supply (+5 V and ±12 V), the mediaconverter and the LP thermostabilization system. One LP can support operation of two clusters. The Central DAQ Post of control and data acquisition ensures operation of the URAN setup in the exposition and monitoring modes, controls all Local Posts of the setup and stores experimental data. The Central DAQ Post includes the central computer, the network equipment, the mediaconverters (Ethernet→Op. Link) and the module of external synchronization. Time synchronization of clusters with an accuracy of 10 ns is performed using GPS/GLONASS systems. At the central computer, information about registered events is processed by special software and is stored in files that contain event timestamps and parameters of the signals from the detectors of triggered clusters. The software also allows automatic control of all elements of the cluster Local Posts of data acquisition.

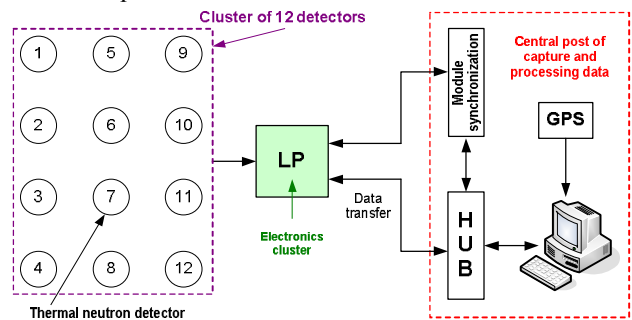

Figure 7: The structure of the cluster of URAN setup.

For the registration of the EAS neutron component, a new type of detector, well-proven in the PRISMA-32 setup (MEPhI, Moscow) [15], is used. The scheme of the detector with the outer enclosure is shown in figure 8.

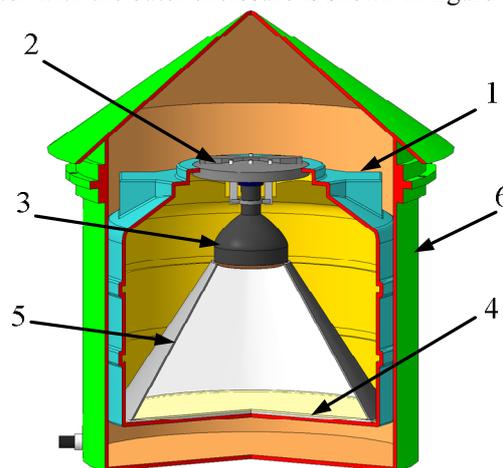

Figure 8: The design of *en*-detector; 1 - light-shielding enclosure of the detector; 2 - a cover with a suspension; 3 - PMT-200; 4 - scintillator ZnS (Ag) + B$_2$O$_3$; 5 - reflective cone; 6 - outer casing (galvanized steel).

The detector housing represents a light-tight black plastic tank with volume of 200 liters (570×740 mm). To improve the light collection, the diffusely reflecting cone is used. The scintillator-compound ZnS(Ag)+B$_2$O$_3$ is located on the base of the cone. The FEU-200 photomultiplier is placed at the top of the cone. The PMT with a high voltage divider mounted on the tank cover using a special holder forms the detecting block. The effective area of the scintillator is $\sim 0.36$ m$^2$.

The detection of thermal neutrons by scintillator ZnS (Ag)+B$_2$O$_3$ occur in the following capture reactions ($\sigma$=3838 barn):

$^{10}$B+n → $^{7}$Li + α + 2.792 MeV
$^{10}$B+n → $^{7}$Li* + α + 2.31 MeV

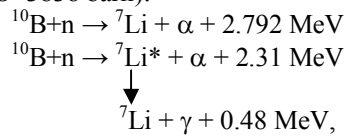

$^{7}$Li + γ + 0.48 MeV,

The scintillator is made in the form of silicon plate with the area of 0.36 m$^2$ and a thickness of 5 mm. The thickness of the scintillator composition is about 50 mg/cm$^2$ and provides thermal neutron detection efficiency of ~ 20%.

## 6. CONCLUSION

Equipping of the experimental complex NEVOD by new unique detectors is aimed at the developing of a new multi-component approach to the study of PCR in a very wide energy interval. Precise studies of the muon component of EAS in a wide range of zenith angles with simultaneous detection of energy deposits of muon bundles in Cherenkov water calorimeter will be conducted for the first time. Such measurements will be complemented by information on the neutron component of EAS by means of the detector also being created for these purposes for the first time.

## Acknowledgments

This work was performed at the Unique Scientific Facility "Experimental complex NEVOD" and was supported by the Ministry of Education and Science of the Russian Federation (contract RFMEFI59114X0002) and MEPhI Academic Excellence Project (contract 02.a03.21.0005, 27.08.2013) and by the grant of the Russian Foundation for Basic Research (project 13-02-12207-ofi-m-2013).

## References

[1] V.M. Aynutdinov et al., "Neutrino Water Detector on the Earth's Surface (NEVOD)", Astrophysics and Space Science, 1998, 258, pp. 105-115.

[2] V.M. Aynutdinov et al., "Detection of muon bundles at large zenith angles" Nucl. Phys. B (Proc. Suppl.), 1999, 75A, pp. 318-320.

[3] N.S. Barbashina et al., "Ultra-high energy cosmic ray investigations by means of EAS muon density measurements", Nucl. Phys. B (Proc. Suppl.), 2007, 165, pp. 317-323.

[4] R.P. Kokoulin et al., "Measurements of the energy deposit of inclined muon bundles in the CWD NEVOD", 2015, J. Phys.: Conf. Ser., 632, p. 012095.

[5] C. Grupen et al., "Cosmic Ray Results from the CosmoALEPH Experiment", Nucl. Phys. B (Proc. Suppl.), 2008, 176, pp. 286-293.

[6] J. Abdallah et al., "Study of multi-muon bundles in cosmic ray showers detected with the DELPHI detector at LEP", Astropat. Phys., 2007, 28, pp. 273-286.

[7] G. Rodriguez et al., "A measurement of the muon number in showers using inclined events detected at the Pierre Auger Observatory", EPJ Web of Conferences, 2013, 53, Article 07003.

[8] V.V. Kindin et al., "Cherenkov Water Detector NEVOD: A New Stage of Development", Phys. Procedia, 2015, 74, pp. 435-441.

[9] V.V. Kindin et al., "Cherenkov Water Detector NEVOD: A New Stage of Development", Phys. Procedia, 2015, 74, pp. 435-441.

[10] A.G. Bogdanov et al., "Investigation of the properties of the flux and interaction of ultrahigh-energy cosmic rays by the method of local-muon-density spectra", Phys. Atomic Nuclei, 2010, 73, 11, pp. 1852-1869.

[11] M.B. Amelchakov et al., "Coordinate detector DECOR for cosmic ray study at large zenith angles", Izv. Akad. Nauk. Ser. Fiz., 2002, 66, pp. 1611-1613

[12] D.V. Chernov et al., "Experimental setup for muon diagnostics of the Earth's atmosphere and magnetosphere (the URAGAN project)", Proc. 29th ICRC (Pune), 2005, Vol. 2, pp. 457-460.

[13] N.S. Barbashina et al., "The URAGAN wide-aperture large-area muon hodoscope", Instrum. Experim. Techn., 2008, 51, 2, p. 180-186.

[14] M.B. Amelchakov et al., "Measuring the spectrum of the local density of charged particles on the SCT setup", Bull. Russ. Acad. Sci.: Phys., 2015, 79, 3, pp. 368-370.

[15] D.M. Gromushkin et al., "The array for EAS neutron component detection", J. Instr., 2014, 9-8, Article C08028

[16] V.V. Borog et al., "Measuring module for registration of Cherenkov radiation in the water", Proc. 16th ICRC (Kyoto), 1979, Vol. 10, p. 380.

[17] http://www.ekran-os.ru/ - JSC "Ekran-optical systems", Novosibirsk, Russia.

[18] G. Battistoni et al., "The NUSEX detector", Nucl. Instrum. Meth. A, 1986, 245 (2-3), pp. 277-290.

[19] V.V. Kindin et al., "Cherenkov water calorimeter on the basis of quasispherical modules", 2015, Proc. of Science, Vol. 30-July-2015, 676.

[20] E.A. Zadeba et al., "Status of a development of the large scale coordinate-tracking setup based on the drift chambers", J. Phys.: Conf. Ser., 2015, 632 012031.

[21] N.I. Bozhko et al., "Drift chamber for the Serpukhov neutrino detector", Nucl. Instrum. Meth. A, 1986, 243, pp. 388-394.

[22] I.I. Yashin et al., "EAS array of the NEVOD Experimental Complex", J. Phys.: Conf. Ser., 2015, 632, 012029.

[23] I.A. Shulzhenko et al., "A proposed NEVOD-EAS installation for the detection of extensive air showers", Bull. Russ. Acad. Sci.: Phys., 2013, 77 (5), pp. 641-643.

[24] M. Aglietta et al., "The EAS-TOP array at Gran Sasso: results of the electromagnetic detector", Nucl. Phys. B (Proc. Suppl.), 1990, 16 (C), pp. 493-494.

[25] A. Chiavassa et al., Proc. 28th ICRC (Tsukuba), 2003, Vol. 2, p. 992.

[26] D.M. Gromushkin et al., "Project of the URAN array for registration of atmospheric neutrons", J. Phys.: Conf. Ser., 2016, 675, 032043.

[27] D.M. Gromushkin et al., "Study of EAS neutron component temporal structure", Astroph. and Space Sciences Transactions, 2011, 7 (2), pp. 115-117.